\documentclass[twocolumn,twoside,slac_two]{revtex4}
\usepackage{graphicx}
\usepackage{fancyhdr}
\usepackage{epsfig} % I added this...

\begin{document}

\title{Prompt VERITAS Observations Triggered by High Energy Fermi-LAT Photons}


\author{D. Staszak, for the VERITAS Collaboration}
\affiliation{McGill University, 3600 rue University, Montreal, Quebec, H3A2T8, Canada}

\begin{abstract}
In any given 24-hour period, the Fermi-LAT detects photons with energies of order 100 GeV not associated with any known VHE emitter.
The RA/Dec of these photons frequently falls into the field of view of VERITAS the following evening.
A single photon in the LAT can potentially mean many detectable photons on the ground due to the superior collection area of the IACT (Imaging Atmospheric Cherenkov Technique) method.
During the 2011-2012 observing season, VERITAS implemented a new target-of-opportunity (ToO) program to trigger VERITAS observations on a small sample of the highest energy LAT photons.
Each photon that triggered observation was detected by the LAT within the previous 24-hour period and was not spatially associated with a known VHE emitter.
We present here the results of these ToO observations, a discussion of the method used, and our future plans.

\end{abstract}

\maketitle

\thispagestyle{fancy}


\section{Introduction}

Since 2008 the Fermi-LAT has supplied the gamma-ray community with a snapshot of the entire sky every 3 hours.
An all-sky search for point sources using an integrated exposure of 24 months of LAT photons yielded over 1,800 point sources \citep{2fgl}.
However, it is likely that beneath the LAT point source detection threshold sit numerous faint gamma-ray sources.
In fact, every day the LAT detects photons with energies of order 100 GeV not associated with any known emitter.
Utilizing timing information to catch flaring or transient events may be one of our best windows to discover the origin of these photons.
While the LAT's survey strategy lends itself well to analyses using long integrated exposures, it leaves the LAT relatively insensitive to 
short-lived moderate flaring behaviour.
We discuss here a novel program to use the LAT all-sky survey to trigger prompt observations of very high energy (VHE: $>$100 GeV) photon positions with VERITAS.
VHE photons provide us with a useful tool since they are detectable both by the LAT and by ground-based imaging Cherenkov detectors.
The collection area of VERITAS is $\sim$$10^{5}$m$^{2}$ larger than that of the Fermi-LAT, so one photon detected by the LAT can lead to many in VERITAS if the photons are in fact originating from a hard source.

Diffuse extragalactic gamma-ray radiation not associated with any known origin (the EGB) was first discovered by the SAS-II instrument
in the 20-200 MeV band \citep{SAS}.
EGRET later found the EGB spectrum to extend up to GeV energies \citep{EGRET1, EGRET2}.
Several groups have now used LAT data to statistically estimate the relative contributions of unresolved point sources (blazars, non-blazar AGN) and truly diffuse emission (starburst/normal galaxies, pulsar systems, DM) in the extragalactic sky.
Taken as a whole, these studies demonstrate a lack of consensus regarding the relative contributions at GeV energies.
%Neronov \& Semikoz\cite{nerSem} and Abazajian et al.\cite{kev} predict that close to 100\% of the extragalactic photons originate from unresolved
\cite{nerSem} and \cite{kev} predict that close to 100\% of the extragalactic photons originate from unresolved
point sources (over the ranges 10 to 400 GeV and 0.01 to 100 GeV, respectively).
%Conversely, Malyshev \& Hogg\cite{hogg} push this number closer to $\sim$20$\%$ over the range of 1 to 300 GeV.
Conversely, \cite{hogg} push this number closer to $\sim$20$\%$ over the range of 1 to 300 GeV.
The latest results from the Fermi-LAT collaboration find between 50 and 80$\%$ of EGB photons to originate from unresolved point sources (over 0.1 to 100 GeV)  \citep{LAT}.
At TeV energies the EGB remains unmeasured.

If the origin of many of these VHE photons is unresolved point sources, as suggested by some, we would expect a sizeable fraction to be BL Lac objects.
BL Lacs have been shown to exhibit a variable gamma-ray flux with a time-dependence ranging from minutes to hours to days.
Our trigger is designed to be as fast a trigger as possible using LAT public data to tell VERITAS where to point. %(VERITAS has a relatively small field of view, $\leq$5.0$^{o}$).
We aim to catch GeV/TeV flaring or transient activity from new VHE emitters.

%Despite the growing list of sources in TeVCat\cite{tevCat}, the TeV energy band is still relatively unexplored.
Despite the growing list of sources in TeVCat\footnote{http://tevcat.uchicago.edu}, the TeV energy band is still relatively unexplored.
We include both galactic and extragalactic latitude photons in this program to make an unbiased search for new sources.
This allows for the possibility of discovering galactic transient events or undiscovered steady-state TeV sources in the galactic plane.

\begin{figure*}
\includegraphics[width=140mm]{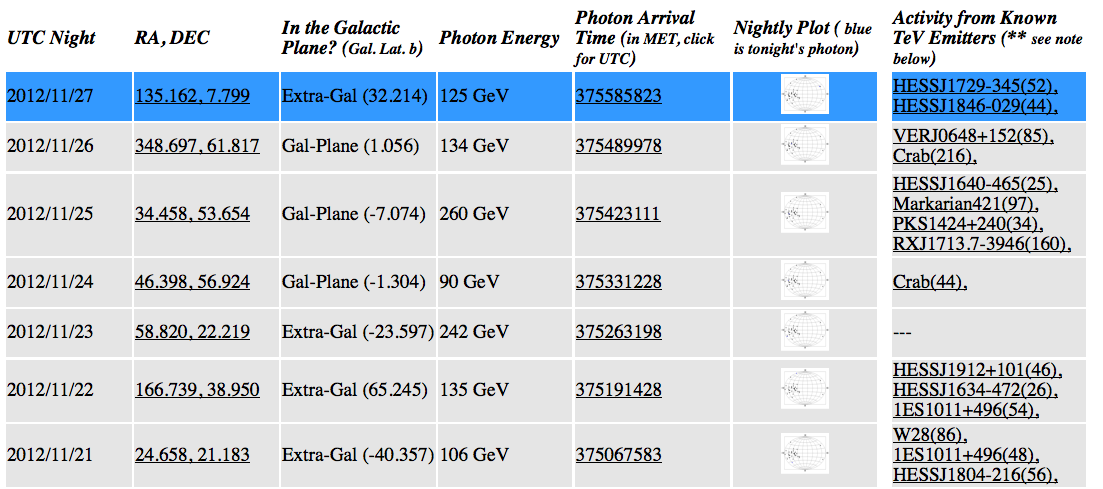}
\caption{Details of the highest energy photon in the view of VERITAS for several nights in 2012.  
Listed for each UTC night is the photon position and energy, whether that photon is within 10$^\circ$ latitude of the 
galactic plane, and photons from the previous 24 hours within 0.5$^{\circ}$ of a known TeV emitter (the energy of that
photon in GeV is given in parentheses).}
\label{screenshot}
\end{figure*}

\section{Algorithm Details}

LAT photon data are typically available for public download within $\sim$8 hours of detection. 
Prior to the start of VERITAS nightly observations, we download the data on all LAT photons from the previous 24-hour period.
These photons are then analyzed using the fermi tools to select high-quality $>$25 GeV 
photons (event class 4, ultraclean).
%We follow the all-sky analysis protocol and use $gtselect$ and $gtmktime$ to specify event and time selection cuts\cite{fssc}. 
We follow the all-sky analysis protocol and use $gtselect$ and $gtmktime$ to specify event and time selection cuts\footnote{http://fermi.gsfc.nasa.gov/ssc}. 
We then filter the remaining photons to remove those spatially associated with known VHE sources
($0.5^{\circ}$ association of a source found in TeVCat)
and find the highest energy photon that rises above $50^{\circ}$ elevation in the view of VERITAS at some point during the night.
A separate algorithm also simultaneously searches these filtered $>$25 GeV photons for two within $0.5^\circ$ of each other in that 
same 24-hour period (a `doublet' trigger).

\section{Triggers}

\begin{figure}
\includegraphics[width=80mm]{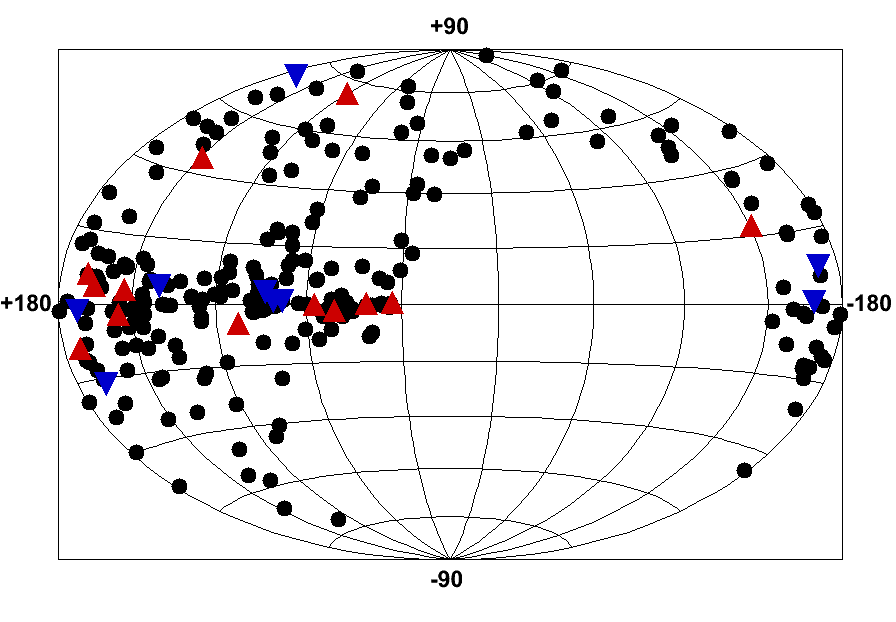}
\caption{Position in galactic coordinates of the highest energy LAT photon in the field of view of VERITAS for each observing night.
Markers in red (up-facing triangles) are those positions which triggered observations and were observed under good weather and with acceptable moon illumination. 
Markers in blue (down-facing triangles) are those positions that were observed in cloudy or poor observing conditions (they may 
or may not have exceeded the trigger threshold energy).}
\label{photonsAitoff}
\end{figure}

Figure \ref{photonsAitoff} shows the position in galactic coordinates of the highest energy LAT photon visible to VERITAS from each 24-hour period of the 2011-2012 observing season.
The black circles represent the full sample of nightly photons and the red triangles (up-facing) represent the subsample of photons that triggered 
VERITAS observation.
Photons with energies exceeding 160 GeV that met the criteria listed in the previous section triggered 20 minutes of observation.
This particular energy threshold was chosen to allow for $\sim$1 trigger per week.

Each trigger was observed for a single 20 minute data run with the possibility of further observation on encouraging signals.
Additionally, each nightly photon was included in a list of `filler' or poor weather targets.  
The IACT detection method is very sensitive to atmospheric and cloud conditions.
Data collected under filler conditions are not typically used in analyses due to large systematic uncertainties and are primarily used to search for flaring behaviour.
Figure \ref{photonsAitoff} also shows photon positions that were observed in filler conditions (blue down-facing triangles).

We triggered good weather observation 14 times for a total of 5 hours.
We also collected filler weather observations on another 9 photon positions for a total of $\sim$5 hours.
Additionally, we observed two doublet triggers in good weather.
The data from each observation were analyzed the following day using two independent analysis packages.
Unfortunately, no significant gamma-ray signals were detected.
Of the two doublet triggers, one was within 0.5$^{\circ}$ of the SNR W44 that is bright in GeV but not detected in the TeV band.
A future publication will present upper limits for each of the triggered photon positions.

\begin{figure}
\includegraphics[width=80mm]{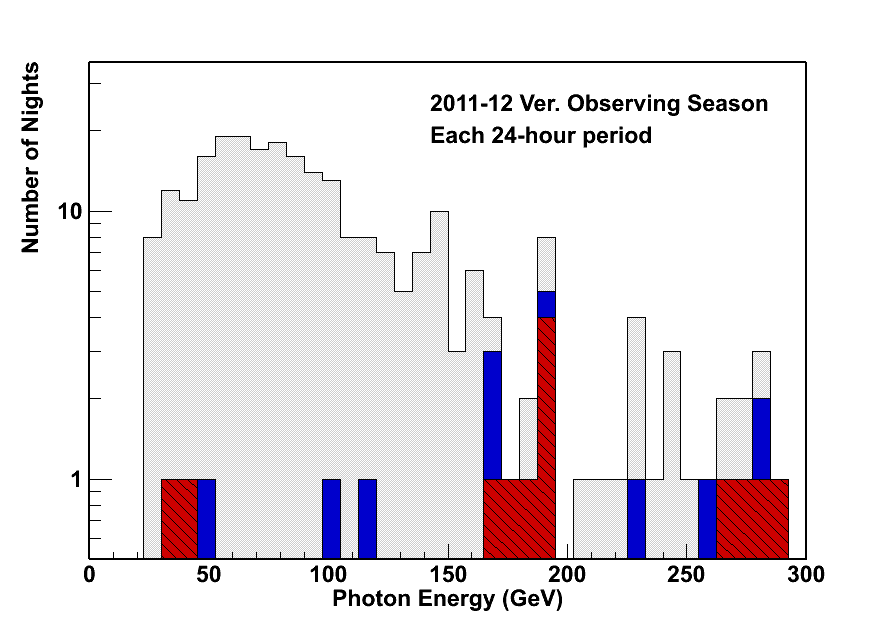}
\caption{Histogram of the highest energy LAT photons in the field of view of VERITAS for each observing night.
The filled blue area represents a subsample of all the photons that were observed (including both good and filler weather observations).
The filled and hashed red area represents a further subsample that triggered observation and were observed in dark, good weather conditions.
The two triggers in red below 50 GeV were doublet triggers.}
\label{photonsFreq}
\end{figure}

\section{Conclusions and Future Plans}

The sensitivity of VERITAS during the 2011-2012 observing season translates 20 minutes of data into a $\sim$10$\%$ Crab Nebula flux detection limit. 
For most known gamma-ray sources this would be an exceptionally strong flare.
All data collected were analyzed and no significant gamma-ray signals were detected.
This is not particularly discouraging since we know $a$ $priori$ that we are searching for dynamic astrophysical objects.
The lower limit of our response time is 8 hours.
Within this window short-lived flares can turn on and off without being detected by VERITAS.

The VERITAS Time Allocation Committee has approved this program for the 2012-2013 
observing season with some adjustments.
Our strategy now is to trigger fewer observations but to take longer exposures on each trigger position (60 minutes vs. 20 minutes). 
To achieve the objective of less frequent triggers we now use 200(225) GeV extragalactic(galactic) thresholds.
The higher relative galactic threshold is used to more evenly smooth our sky exposure.
These longer exposures will improve our detection limit to the $\sim$5$\%$ Crab Nebula flux level.
Additionally, new higher quantum efficiency PMTs were installed in each of the four 
VERITAS cameras during the summer of 2012.
We expect a further boost in our sensitivity from this upgrade since the energy threshold of the array has been substantially lowered \citep{kieda}.

We report results from a ToO program triggering VERITAS observations on VHE LAT photons from locations without known emitters.
We utilize the daily full-sky snapshots that the LAT provides to guide VERITAS where to look.
The time used to run this program is relatively small but the discovery potential is huge; one discovery using this technique can potentially give valuable input about the origin 
of the EGB.
We note that our methods can be easily adapted to find and report high-energy LAT photons visible by other observatories, if there is interest.
We are open to the use of this program by other collaborations and encourage participation by H.E.S.S. and MAGIC.

\bigskip % extra skip inserted
\begin{acknowledgments}
This research is supported by grants from the U.S. Department of Energy Office of Science, the U.S. National Science Foundation and the Smithsonian Institution, by NSERC in Canada, by Science Foundation Ireland (SFI 10/RFP/AST2748) and by STFC in the U.K. We acknowledge the excellent work of the technical support staff at the Fred Lawrence Whipple Observatory and at the collaborating institutions in the construction and operation of the instrument.

\end{acknowledgments}

\bigskip % extra skip inserted

\end{document}